\documentclass[a4paper,10pt]{article}
\usepackage[utf8x]{inputenc}

\usepackage[pdftex]{graphicx}	
\usepackage{hyperref}							
\usepackage{amsmath, amsthm, amssymb}
\usepackage{color}
\usepackage{epsfig}
\usepackage{verbatim} % for commenting sections
\usepackage{epstopdf}
\usepackage{caption}
\usepackage{subcaption}
\usepackage{booktabs}% for tabulars

\usepackage[hmargin=3cm]{geometry}

\title{Heterogeneous Enterprises in a Macroeconomic Agent-Based Model}
\author{Cornelia Metzig$^\dagger$ and Mirta B. Gordon$^\dagger$,$^*$ \\
$^\dagger$Universit\'e Joseph Fourier Grenoble 1 / $^*$CNRS\\ Laboratoire LIG \\
\texttt{cornelia.metzig@imag.fr}, \texttt{mirta.gordon@imag.fr}
}
\date{}
\begin{document}

\maketitle
%\newpage
%\tableofcontents
%\newpage

\begin{abstract}
 We present a macroeconomic agent-based model that combines several mechanisms operating at the same timescale, while remaining  mathematically tractable. It comprises enterprises and workers who 
 compete in a job market and a commodity goods market. The model is stock-flow consistent; a bank lends money charging interest rates, and keeps track of equities. Important features of the model are heterogeneity of enterprises, existence of bankruptcies and creation of new enterprises, as well as productivity increase. The model's evolution reproduces empirically  found regularities for firm size and growth rate distributions. It combines probabilistic elements and deterministic dynamics, with  relative weights that may be modified according to the considered 
problem or the belief of the modeler. 
We discuss statistical regularities on enterprises, the origin and the amplitude of endogeneous fluctuations of the system's steady state, as well as the role of the interest rate and the credit volume. We also summarize obtained results which are not discussed in detail in this paper. 
\end{abstract}

\section{Introduction}

 Macroeconomic agent-based models are a relatively young strand of research \cite{DelliGatti2011}, \cite{Bruun2008}, \cite{Seppecher2009}, \cite{DelliGatti2008}. They allow to propose microscopic underlying mechanisms that lead to empirically observed statistical regularities of a macroeconomy, which models starting directly on the aggregate level cannot provide. They avoid the so-called ``fallacy of composition``, since they allow to study interactions. Macroeconomic agent-based models are often conserving the flow of funds, i.e. they can serve as a tool to study the role of debt levels, interest flows, or availability of credit on economic activity. The most important implication of this stock-flow consistency is that the financial condition of agents plays a central role determining their actions. In that sense, stock-flow consistent models provide an alternative to efficient markets as a guiding principle of an economic model, where ``the impact of the flow of funds and the stocks of credit and debt are fully reflected in returns and risks at the individual level''\cite{Bezemer2010}. If, in contrast, agents base their decisions only on their balance sheet, and do not know the financial condition of others, systemic risk can be present. %\textcolor{red}{Aware of this simplification, this paper investigates which are the phenomena that can readily be explained with it.}

Stock-flow consistent models use as main principle the conservation of money. They can be agent-based, such as \cite{Seppecher2009},\cite{Bruun2008} and this model, or analytical. Several analytical stock-flow consistent models describe the money flows between sectors \cite{Keen2010}, \cite{GodleyBook2007}. They exhibit steady state scenarios where the money flows to and from each sector add to zero. This implies that, if the model is discrete in time, in one iteration every actor needs to re-inject in the system the money he has received, be it wages, profits or interest payments. If this does not happen, economic activity will eventually cease. Regarding enterprises, this phenomenon has become known as the `paradox of monetary profits' \cite{BruunHeynJohnsen2008}, originally formulated by Marx, stating that enterprises can at most earn what they have paid in wages. To circumvent this, other models introduce as money flows interest payments of enterprises to the bank, which is a joint stock company and distributes a dividend \cite{Seppecher2009}, consumption of the bank \cite{Keen2010}, or dividends of enterprises who are joint-stock companies \cite{Bruun2008}. The stock-flow consistent model proposed here avoids the paradox of monetary profits because enterprises spend a part of their gross profits to pay interests on debts to the bank, and the rest in the goods market. The bank who is the recipient of the interest payments will bear losses from loans that are not reimbursed whenever an enterprise goes bankrupt. As the simulations show, in the latter case, money flows do not equal within one iteration, since the accumulation of debts goes over many iterations. On the long term, each sector will re-inject what he has received, but this steady state exhibits fluctuations whose strength depends on the parameters of the system.

 A major difference to analytical stock-flow consistent models is that agent-based macroeconomic models allow for the description of competition all markets. \cite{DelliGatti2011} and \cite{Seppecher2009} use a price mechanism in the goods market, the job market and the credit market. In the model proposed in this paper, competition of enterprises does not happen via prices but due to limited quantities. Enterprises compete for the available amount of workers in the job market, and for aggregate demand in the goods market respectively. These market allocations introduce a stochastic term in the dynamics of the system. In order to be able describe the arising stochastic model theoretically, it needs to be simpler than the scenario presented in this paper.

%In that aspect the model can be seen in the widely studied context of stochastic processes \textcolor{red}{citations here ? \cite{Takayasu1997}, \cite{Marsili1998}, \cite{Zanette1997}, \cite{BlankSolomon2000}}, and be compared to models focusing only on a single stylized fact such as \cite{Botazzi2006} or \cite{Gibrat1931}.
%Theoretically, a simpler setting of the model (without interest payments, without entry and exit and without heterogeneous parameters of enterprises) is analyzed in a forthcoming paper \cite{MetzigGordon2012Physica}, in which we provide explanations for a power law size distribution, and a tent-shape growth rate distribution and most importantly, the link between these two. 

%Nevertheless, the usefulness of the simple setting analyzed in \cite{MetzigGordon2012Physica} may be questioned. 
In his paper on the history of growth models, J. Sutton revealed two aspects limiting the use of such simple models \cite{Sutton1997}: In the past many purely stochastic models had been presented, which do not attribute any importance to the purposeful behaviour of agents. This has been seen as a flaw by many economists, who turned towards game-theoretic approaches. Secondly, it can be difficult to interpret models with very general assumptions. This paper takes a different approach: it combines the well-understood stochastic model with economically relevant features such as margin heterogeneity of enterprises, credit constraints, exit of enterprises through bankruptcy, creation of new enterprises, and productivity increase. These features are in fact complementary and counter-acting mechanisms, implying that a steady state of the system can be found in which the system exhibits statistical regularities. Along with the introduction of these features comes the difficulty -- omnipresent in macroeconomic agent-based models -- to weight the dominance of the mechanisms via the choice of the parameters. Not only different empirical datasets, but also different viewpoints might lead to different choices.   

The model studies the coexistence of two  mechanisms: on the one hand, a mechanism introducing a random growth term due to the competition for limited quantities, which on its own yields an explanation for size distribution and growth rate distribution. On the other hand, a mechanism comprising growth via the creation of new more profitable enterprises, as well as interest payments and bankruptcies. We analyze parameter dependencies of the behaviour both on firm level and on aggregate level. The results of the model are consistent with a number of empirical studies, but validity of this model could be verified with a multivariate database on firms, and possibly extended.

The paper is organized as follows. In section \ref{sec:model}, we introduce the model. In section \ref{sec:results} we show and discuss numerical results, in section \ref{sec:validation} we compare the obtained to existing empirical studies. Finally, in section \ref{sec:conclusion}, we conclude and point at possible extensions and applications.

\section{The model} %%- equations
\label{sec:model}
We present here the main elements of the model, whose workings will be introduced successively in the following sections. We consider three kinds of agents, namely enterprises, workers and a bank. Each enterprise is characterized by its margin $\mu_i$. They all produce the same type of good that is put in the market at the same price, $p$. This commodity good is an abstraction of purchases in the real sector. It is a useful concept to ensure that via limited aggregate demand all enterprises are in competition, since the decisions of enterprises are governed by their profits. Enterprises compete only in two quantities, for sales (i.e. the aggregate demand) and for workforce. 
The number of hired workers, that depends on the enterprise profits and workforce availability, allows us to characterize the enterprise size. All workers receive the same wage $w$ per period when hired, while unemployed workers do not get anything. Enterprises and workers spend their earnings in the goods market. The bank may lend money to the enterprises, mainly to pay wages, and determines the corresponding interest rate. Enterprises with large liabilities may go bankrupt, and a constant flow of young enterprises more or less compensates for these disappearances.

\subsection{Main elements}
More precisely the system is composed of $N_e$ enterprises (whose number may fluctuate along the iterations, according to the corresponding setting), $N_w$ workers and the bank: 

\begin{description}

\item [enterprises] $i$ ($1 \le i \le N_e$) are characterized by an expected gross margin $\mu_i \equiv$ (expected sales - wage expenses)/expected sales, that we call just {\em margin} hereafter. 
They produce $q_i$ non-durable goods per period, that are put on the market at a price $p$. %If all the  $q_i$ produced goods were effectively sold\footnote{In section ... we present more in details the difference between expected and actual quantities.},
The margin writes as follows: %%\textcolor{red}{the gross margin is always defined like that, independently of the actual sales!}
\begin{equation}
\mu_i = \frac{p \, q_i -n_i\, w} {p \, q_i}\,,
\label{eq:margin}
\end{equation}
where $n_i$ is the number of hired workers. In this paper we do not allow enterprises modify their margins: $\mu_i$ are intrinsic parameters drawn at random with a uniform probability density in the range $[\mu_m \leq \mu_i \leq \mu_M ]$ with $0 < \mu_m < \mu_M < 1$.
Notice that these expected gross margins are distinct from the net margin actually earned by enterprises, detailed in equation (\ref{eq:munet}), which may be negative. The margin (\ref{eq:margin}) reflects the productivity and the technological level of the enterprise: the higher the margin, the lower the production cost per unit or the lower the number of workers needed to produce a given quantity.

\item[workers] $j$ ($1 \le j \le N_w$) may be employed or unemployed. When employed they earn a wage $w$ per period. Workers do not save, unless they cannot satisfy their demand for goods: like Godley and Lavoie~\cite{GodleyBook2007} in their `simplest model', and M. Kalecki~ \cite{Kalecki1942} in his profit equation, we assume that workers try to spend all their earnings in the consumption market.

\item [the bank] lends money to enterprises at a constant interest rate $r$, and keeps track of their equities, which cannot be less than a lower bound $u_i \le 0$ that depends on the enterprises' wage bills:
\begin{equation}
u_i = - \gamma \, w \, n_{i}\,,
\label{bankrthreshold}
\end{equation}
where $\gamma$ is some positive constant. Enterprises that do not satisfy this constraint are declared bankrupt. 
\end{description}

\begin{figure}
	\centering
		\includegraphics[angle=0, width =0.95\textwidth]{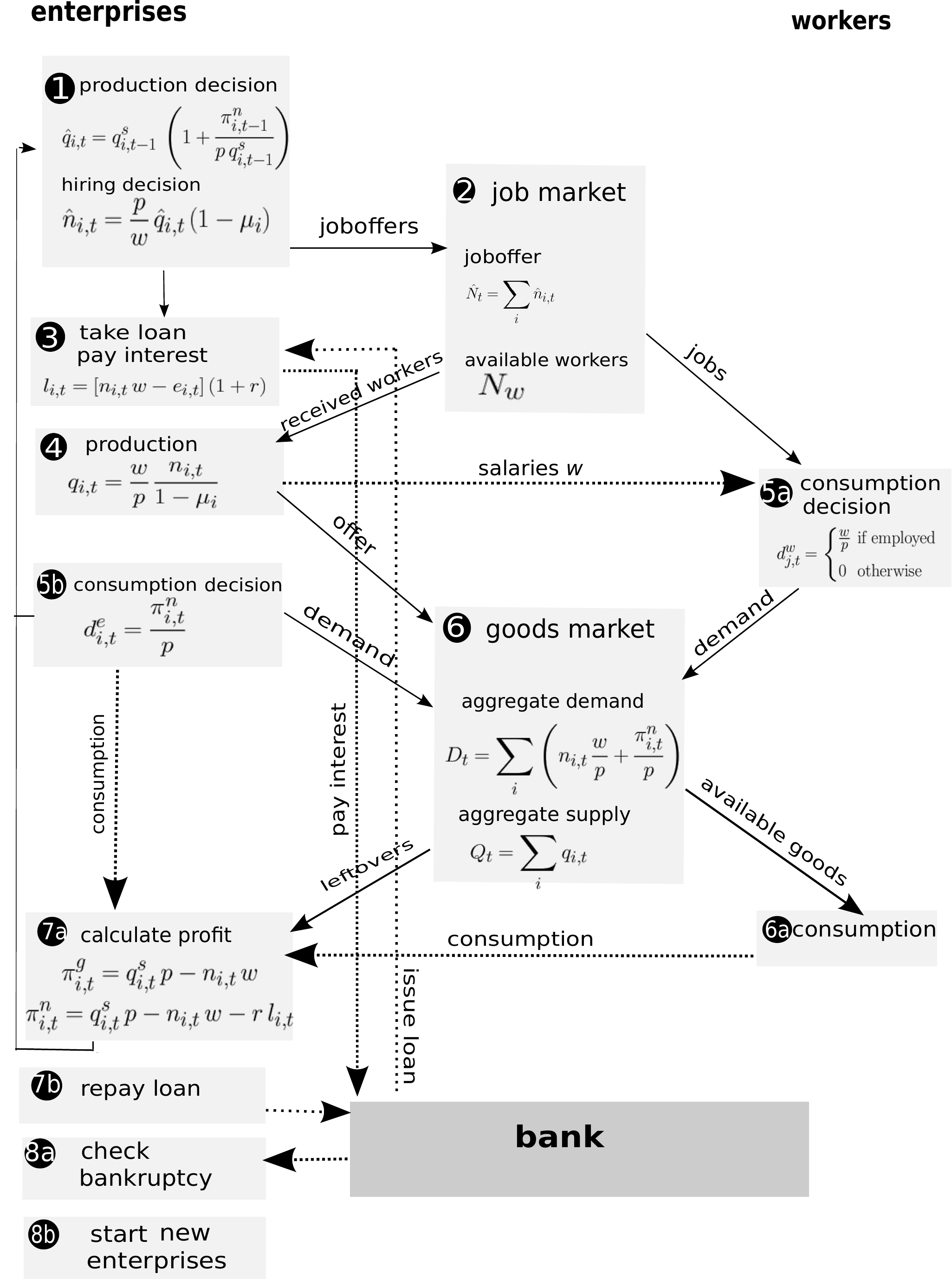}
 	\caption{Scheme of the model. The arrows denote interactions, dotted lines denote money flows. The numeration corresponds to the numeration in the description.}
	\label{fig:scheme_complete}
\end{figure}

\subsection{Dynamics}
The dynamics of the model, visualized in figure \ref{fig:scheme_complete}, is detailed below. We use upper case letters for aggregate quantities, lower case letters for individual quantities, both for firms $i$ or workers $j$. Quantities with a hat ($\hat q_i$, $\hat n_i$) are initially planned quantities intended by firms before checking constraints.

{\bf at each time step t:}

\begin{enumerate}
\item \textbf{production decision of enterprises:} enterprises calculate their expected production volume $\hat q_{i,t}$ according to their previous iteration sales $q^s_{i,t-1}$ and net profits $\pi^n_{i, t-1}$ (from which interest payments are already subtracted, see equation (\ref{eq:pi_net})). If profits are positive (negative) they plan to produce more (less) goods proportionally to their net profit ratio per sold unit:
\begin{equation}
\hat q_{i,t}= q^s_{i, t-1}\, \left(1+\frac{\pi^n_{i, t-1}}{p\,q^s_{i, t-1}}\right)\,.
\label{eq:q_p_i}
\end{equation}

%%\item \textbf{Hiring decision.} 
The number of workers necessary to produce the $\hat q_{i, t}$ goods is determined by the intrinsic margin defined in equation (\ref{eq:margin})

\begin{equation}
\hat n_{i,t}=\hat q_{i,t}\,(1-\mu_i) \,\frac{p}{w}.
\label{eq:n_p_i}
\end{equation}

\item \textbf{job market:} the aggregate job offers of enterprises 
\begin{equation}
\hat N_{w,t} \equiv \sum_i \hat n_{i,t}
\end{equation}
may exceed or fall behind the available number of workers $N_w$. In our simple job market, workers do not stay at their employer. They are placed anew at each iteration through a process where job offers and demands are matched at random. If $\hat N_{w,t} \leq N_w$ every worker has the same chance of being hired by some firm, but $N_w - \hat N_{w,t}$ workers are left unemployed ; if $\hat N_{w,t} > N_w$, each job opening has the same probability of being filled, there is full employment and $\hat N_{w,t} - N_w$ positions remain vacant. Depending on availability each firm $i$ will hire a number of workers 
\begin{eqnarray}
n_{i,t}=\hat n_{i,t}\text{ if } \hat N_{w,t} \leq N_w\\
\left\langle n_{i,t}\right\rangle=\hat n_{i,t}\,\frac{N_w}{\hat N_{w,t} } \text{ if } \hat N_{w,t}  > N_w\,.
\end{eqnarray}

It is useful to define an \textit{effective $\mu$} of the system, which is the average $\mu$ at which the workers are hired: 
\begin{equation}
 \mu_{eff, t}=\frac{1}{N_{w}}\sum_{i}n_{i,t}\, \mu_i.
\label{eq:mu_eff}
\end{equation}
Its time evolution will allow us to characterize the dynamics of the system. 

\item \textbf{credit market:} once the number of hired workers $n_{i,t}$ is known, each enterprise calculates whether its owned equities $e_{i,t}$ are sufficient to pay the corresponding wages. If necessary, it takes out a loan of amount $l_{i,t}$ from the bank. The latter determines the interest rate $r$, which in the following is assumed to be time independent and the same for all the enterprises. The borrowed amount takes into account the interests to be included upon repayment of the debt:
\begin{equation}
l_{i,t}=[n_{i,t}\, w-e_{i,t}]\,(1+r)
\label{eq:loan}
\end{equation}

%\item \textbf{Payment of wages.} Hired workers receive a wage $\mathbf{w}$. In this basic setting unemployed workers do not receive anything. 
\item \textbf{production:} the amount of goods $q_{i,t}$ actually produced by the $n_{i,t}$ workers hired by firm $i$ (and put in the market) is given by equation:
\begin{equation}
q_{i,t}=\frac{w}{p}\frac{n_{i,t}}{1-\mu_i}\,,
\label{eq:q_o_i}
\end{equation}
so firms with a higher $\mu_i$ produce more goods per worker. The aggregate output $Q_t$ is
\begin{equation}
 Q_t=\sum_{i}q_{i,t}.
\label{eq:GoodsOffer}
\end{equation}
%(Output of an enterprise in monetary units is $p\,q_{i,t}$.) 
%\textcolor{blue}{Is this necessary? Because the true monetary value of the output is the cost.}

\item \textbf{consumption decision:}  workers $j$ intend to spend their wages, which allow them to buy a quantity given by:
\begin{equation}
d_{j,t} =\begin{cases}\frac{w}{p}&\text{if employed}\\
0&\text{otherwise}\end{cases}
\end{equation}
Firms $i$ intend to spend their expected profits:
\begin{equation}
 d_{i,t}=\frac{\pi_{i,t}^n}{p}
\end{equation}

The bank does not act as a consumer (as is done e.g. in \cite{Keen2010}), it can merely lose money when enterprises go bankrupt, which compensates for its interest revenues. Thus, the aggregate demand is:
\begin{equation}
%% D_t=\sum_i (n_{i,t}\,w+\alpha_{i,t}\pi_{i,t})\,.
D_t = \sum_j d_{j,t} + \sum_i d_{i,t}.
\label{eq:demand_all}
\end{equation}

\item \textbf{commodity goods market:} %%\textcolor{red}{I actually would prefer to leave this in two sections because like that in the markets is purely the matching, and otherwise production also could be inside}%\textcolor{blue}{I would merge this point with the previous one, under the title: commodity goods market. \\} 
in the case of excess production, the overall offer (\ref{eq:GoodsOffer}) exceeds aggregate demand (\ref{eq:demand_all}), i.e. $Q_t > D_t$. We assume that each commodity good has the same chance of being sold. This is the situation mainly considered in the present paper. If $D_t > Q_t$, all the production is sold leaving some demands unsatisfied. Each demand -- whether coming from enterprises or workers -- has the same chance of being fulfilled and the agents (workers or enterprises) may save and eventually spend their savings in the next iteration. (However, in the simulations presented in this paper holds $Q_t > D_t$). 
Notice that $Q_t$ and $D_t$ need not be known to the enterprises or workers; they experience if their offer and demand are satisfied by a matching algorithm. %In our simulations a matching algorithm allows us to take care of this constraint. 
%%\textcolor{red}{In the cases presented in this model, aggregate demand is the constraint of the system, so it never happens that a demand is not satisfied, it is the offer that is not satisfied, that is why I inverted the order of the two}. 
Sales of enterprise $i$ are
\begin{eqnarray}
q^s_{i,t}= q_{i,t}\text{ if } Q_t < D_t \\
\left\langle q^s_{i,t}\right\rangle= q_{i,t} \, \frac{D_{t}}{Q_t}\text{ if } Q_t > D_t\,,
\label{eq:q_sold}
\end{eqnarray}

 In the second case where enterprises face a random constraint due to limited aggregate demand, unsold goods are lost. In that case the realized gross margin (see eq. (\ref{eq:pi_gross}) may be lower than the intrinsic margin $\mu_i$, and even be negative, as is the case in the examples shown in this paper in figure \ref{fig:trajectories_mu}. (Sales in monetary units are $p\, q^s_{i,t}$.)

\item \textbf{enterprises balance sheet accounting:} %%\textcolor{red}{(workers are mentioned already above)} 
 firms must pay interests on the whole amount of their loans, i.e. current lending and accumulated old debts. If they have enough assets, they repay the loans at the end of the iteration; if the liabilities exceed the assets, they start the next iteration with a negative net equity. 

In order to prepare the production decision of the next period, firms calculate their gross and net realized profit $\pi^g_{i,t}$ and  $\pi^n_{i,t}$ (which are identical if $r$=0), defined respectively by
\begin{eqnarray}
\pi^g_{i,t}=q^s_{i,t}\,p- n_{i,t}\,w , \label{eq:pi_gross}\\
\pi^n_{i,t}=q^s_{i,t}\,p- n_{i,t}\,w- l_{i,t}\,r. \label{eq:pi_net}
\end{eqnarray}
Notice that the normalized gross profit $\pi^g_{i,t}/(q^s_{i,t}\,p)$ is \textit{not} necessarily identical to the margin $\mu_i$, which is only reached if all the produced goods are sold. %In the net 
The gross and net realized margins of an enterprise %\textcolor{blue}{It would be better to avoid "net realized margin", one can call it normalized net profit. The normalized gross profit can also be negative.}
\begin{eqnarray}
\mu_{i,t}^g=\pi^g_{i,t}/(q^s_{i,t}\,p)\,, \label{eq:mugross}\\
\mu_{i,t}^n=\pi^n_{i,t}/(q^s_{i,t}\,p) \label{eq:munet}
\end{eqnarray}
are $\leq \mu_i$ and can both take negative values. 
 
\item \textbf{bankruptcy and new enterprises:} enterprises whose equities after repayment of loans (as far as repayment is possible) fall below the threshold defined by eq. (\ref{bankrthreshold}) are declared bankrupt, and are removed from the system. Their (negative) equities % \textcolor{blue}{ or their negative financial balance, or negative equities?} 
are losses to the bank. 

In order to avoid a decline in the number of enterprises due to bankruptcies, we introduce a small number of new enterprises at each iteration. Notice that this differs from the approach by Bruun \cite{Bruun2008} and DelliGatti et al. \cite{DelliGatti2011}, who both replace systematically each bankrupt enterprise by a new one, keeping thus the total number of enterprises constant. In our model new enterprises are started at a constant rate $\nu$ per iteration, 
with an initial number of workers ($n_{i,init}$) drawn at random between 1 and 3, and a margin $\mu_i$ drawn at random within the same range of margins as the initial enterprises. With this procedure the total number of enterprises fluctuates upon time. 

In order to endow new enterprises with the possibility to grow faster than existing enterprises and thus slowly displace them, the margin of existing enterprises is recentered at each iteration before new firms are started, such that the new enterprises enter a system with the same $\mu_{eff}$ throughout the simulation: 
%\textcolor{red}{No, in your equation the $\mu_{eff,t}$ and $\mu_{eff,t-1}$ had the same sign so both would be subreacted, it is as follows}
\begin{equation}
 \mu_{i,t}=\mu_{i,t-1}-(\mu_{eff,t} - \mu_{eff,t-1}),
\end{equation}
$\mu_{eff}$ has a tendency to grow, since according to equation \ref{eq:n_p_i} workers tend to move towards enterprises with a higher margin. The recentering is thus a subtraction and its effect is that the $\mu_i$ of existing enterprises slowly decline, like an ageing effect; new enterprises may grow fast because they may have larger margins than the older ones.
This recentering is important for the dynamics of the system: without it, new enterprises would need to start with a higher and higher margin in order to be able to influence the dynamics of the system and displace old enterprises. Then, $\mu_{eff}$ would continue to rise, which is undesired, since $\mu_{eff}$ should remain in a (more or less) constant relation to the interest rate, such that the interest payments represent the same burden for enterprises throughout the simulation. %If in contrast new enterprises are assigned a margin in the same range as existing ones, but no recentering of old enterprieses' $\mu_i$ is performed, new firms would drown within the existing enterprises, and big old enterprises would dominate.

%% \item \textbf{restart of enterprises:} Since the number of active enterprises would decline in the presence if interest payments and a bankruptcy threshold, new enterprises are restarted. Contrarily
\end{enumerate}

\subsection{The role of heterogeneity and financial constraints in the model}
\label{sec:heterogen}

Heterogeneity of $\mu_i$ implies that due to equation (\ref{eq:n_p_i}) enterprises with a higher $\mu_i$ will offer more posts and therefore grow faster than enterprises with low $\mu_i$. On its own, this dynamics would at first form a power law for enterprise sizes, but on the long term converge to a monopoly of the enterprise with the highest margin. Therefore, $\mu$-heterogeneity needs a counterbalance, constisting in this model of the renormalization of the margin, interest payments, bankruptcies and the creation of new enterprises. These mechanisms allow to describe statistical regularities rather than monopoly formation.
%We discuss the effects of heterogeneity of enterprises, credit constraints, productivity increase,  exit and entry of new enterprises. 

\section{Results}
\label{sec:results}
In this section we discuss results of simulations corresponding to different sets of parameters of the model: 
\begin{itemize}
%\item $N_w$, the available workforce (number of workers),
%\item $N_e$, the initial number of enterprises,
\item $[\mu_m,\mu_M]$, the range of expected margins,
\item $r$, the interest rate,
\item $\gamma$, the bankruptcy threshold,
\item $\nu$, the number of new enterprises entering per iteration.
\end{itemize}
In all presented settings enterprises compete for demand, whenever unemployment drops to zero, they compete in addition for workforce (which is not always the case).
The wages are fixed $w=30\, p$ in units of the price of the goods, that are the monetary unit in our simulations. %(but since the value of $w$ is a constant for all enterprises, it does not play any role for the dynamics of the system). 
Given $N_w$, a number of initially active enterprises $N_e$ is chosen, and their margins $\mu_i$ are drawn at random within a given interval  $[\mu_m,\mu_M]$. The number of their employees is chosen respecting some level of initial unemployment, and the desired initial size distribution. The sum of money in the system is zero. Enterprises start with $e_{i,t}\approx 0$ (except a small random value for debts such that the bankruptcies of initial enterprises do not occur in a very small timespan). This initialisation does impact the trajectory of the system, but for the analysis the time long after the start is interesting, where the system found its steady state and is dominated by the choice of $[\mu_m,\mu_M]$, $r$, $\gamma$ and $\nu$.

%% The system has been simulated for many different parameter sets. The results are discussed in the following and examples shown.
\subsection{Steady states and parameter dependence}

\begin{figure}[h!]
	\centering
\begin{subfigure}[b]{0.45\textwidth}
\includegraphics[angle=270, width=\textwidth]{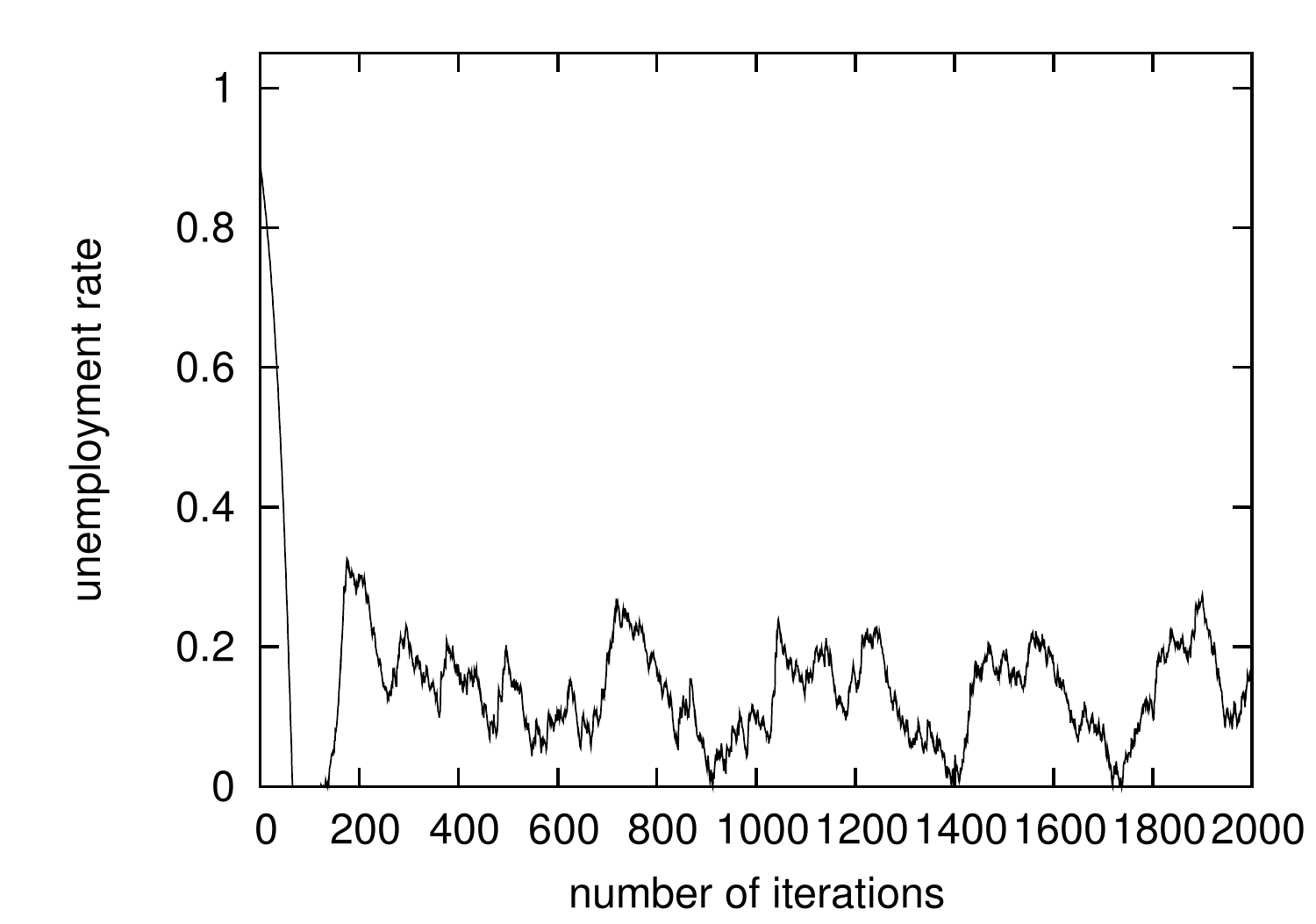}
\caption{}
\end{subfigure}
%\begin{subfigure}[b]{0.45\textwidth}
%\includegraphics[angle=270, width=\textwidth]{im330b/flow_sim330b_2_8.pdf}
%\caption{}
%\label{fig:330b_unempoyment}
%\end{subfigure}
\begin{subfigure}[b]{0.45\textwidth}
\includegraphics[angle=270, width =\textwidth]{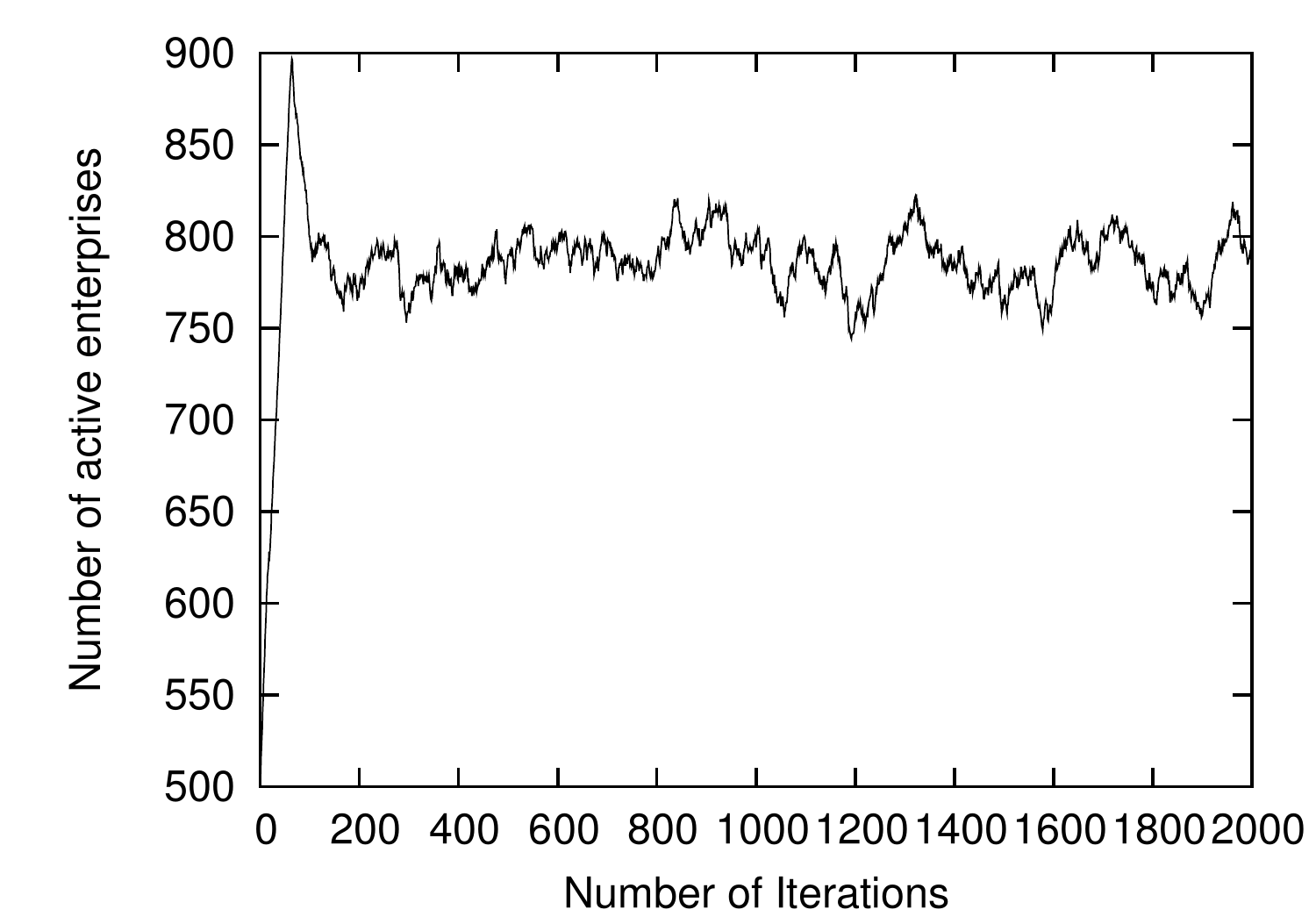}
%\label{fig:330b_number_enterprises}
\caption{}
\end{subfigure}
\begin{subfigure}[b]{0.45\textwidth}
\includegraphics[angle=270, width =\textwidth]{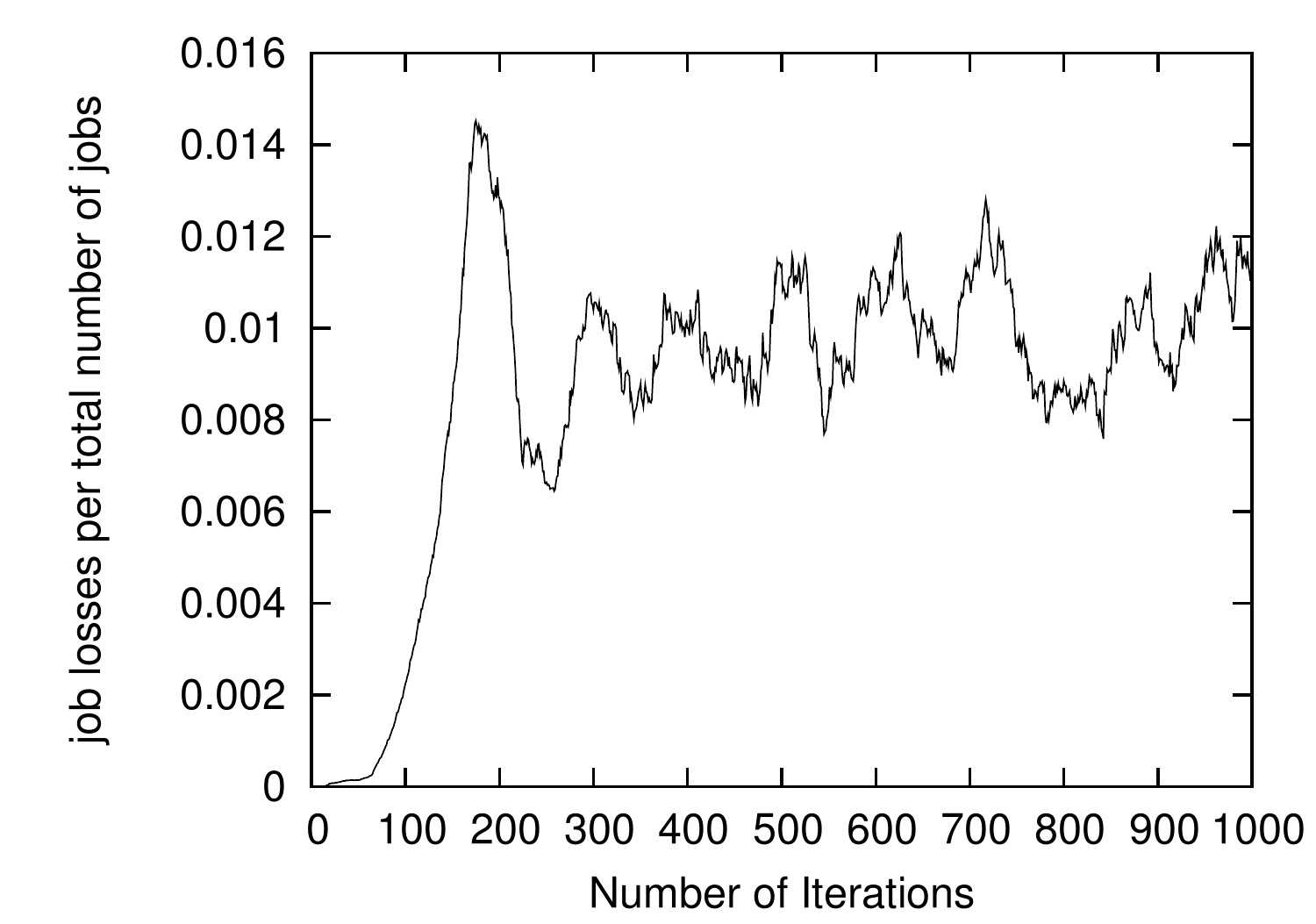}	
%\label{fig:330b_job_losses}
\caption{}
\end{subfigure}
	%\label{fig:sim330b}
 	\caption{$N_w=10\,000$, bankruptcy threshold $\gamma=2$, interest rate $r$=0.011, new enterprises per iteration $\nu=8$, $[\mu_m,\mu_M] = [0, 0.1]$. 
	Time evolution of (a) unemployment, %(b) money flows, 
(b) number of active enterprises, (c) job losses due to firm bankruptcies (first 1000 iterations, averaged over 50 iterations). 
	}
	\label{fig:sim330b}
\end{figure}

Figure \ref{fig:sim330b} presents results corresponding to a typical time evolution of the system. Even after 2000 iterations there are fluctuations due to the randomness in the system. Job losses do not occur at a constant rate but fluctuate around a certain rate (here 1\% of the existing jobs in one iteration. In subfigure (c) of figure \ref{fig:sim330b} job losses are averaged over 50 iterations in order to see slower trends, the actual fluctuations are larger).

The system eventually reaches a steady state, characterized by a number of active enterprises and a level of unemployment, which depends on the interest rate $r$ of the system, the bankruptcy threshold $\gamma$, the number $\nu$ of entering enterprises per iteration and the rannge $[\mu_m,\mu_M]$. The higher the interest rate, the faster the bankruptcy threshold is reached by indebted enterprises, and
the higher needs to be the number of new entries per iteration in order to maintain the same level of employment. The steady state is noisy since the dynamics has two stochastic elements, but for a given set of parameters the system converges to such a state for any initial level of unemployment, any initial size distribution of enterprises, and (almost) any number of initially active enterprises. 

The system fluctuates in the steady state for the following two reasons: firstly, since enterprises spend a part of their profits for interest payments, they cannot spend all their gross profit in the goods market. Thus, their contribution to the aggregate demand falls short  of the one required to absorb the offer. As a result 
enterprises are not able to sell their entire production. The profits of enterprises fluctuate due to the random allocations of demand to every offered good, and so does their production decision in the next iteration. Notice that in the case of full employment, the same random allocation holds additionally for the allocation of workers in the job market. The second reason for fluctuations is due to the random initialization of new enterprises' margins $\mu_i$: depending on its value they grow, decline and incur debts at different speeds, so the job losses due to bankruptcies fluctuate over time. The system evolution keeps memory of the past trajectory through the debt distribution among enterprises. Large job losses due to bankruptcies occur in waves, causing lower demand and therefore on average lower net realized profits for firms. This also impacts the chances and speed of growth of a new enterprise endowed with a higher net margin. Fluctuations are discussed more in detail in section \ref{sec:fluct}.

\subsection{Lifetime of enterprises}
\begin{figure}[h!]
	\centering
\begin{subfigure}[b]{0.45\textwidth}
\includegraphics[angle=270, width =\textwidth]{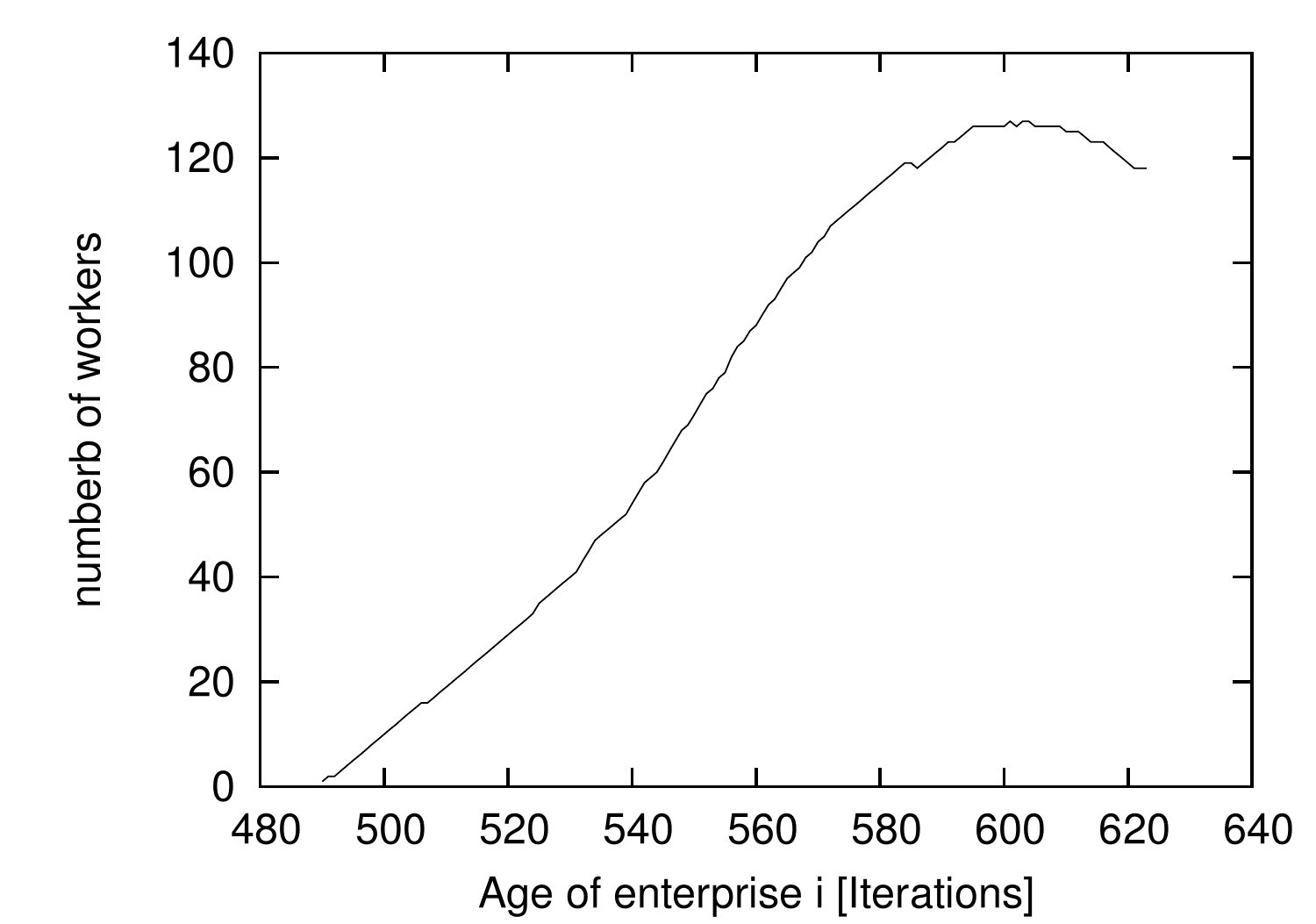}
\label{fig:333/example_hir17}
\caption{}
\end{subfigure}
\begin{subfigure}[b]{0.45\textwidth}
\includegraphics[angle=270, width =\textwidth]{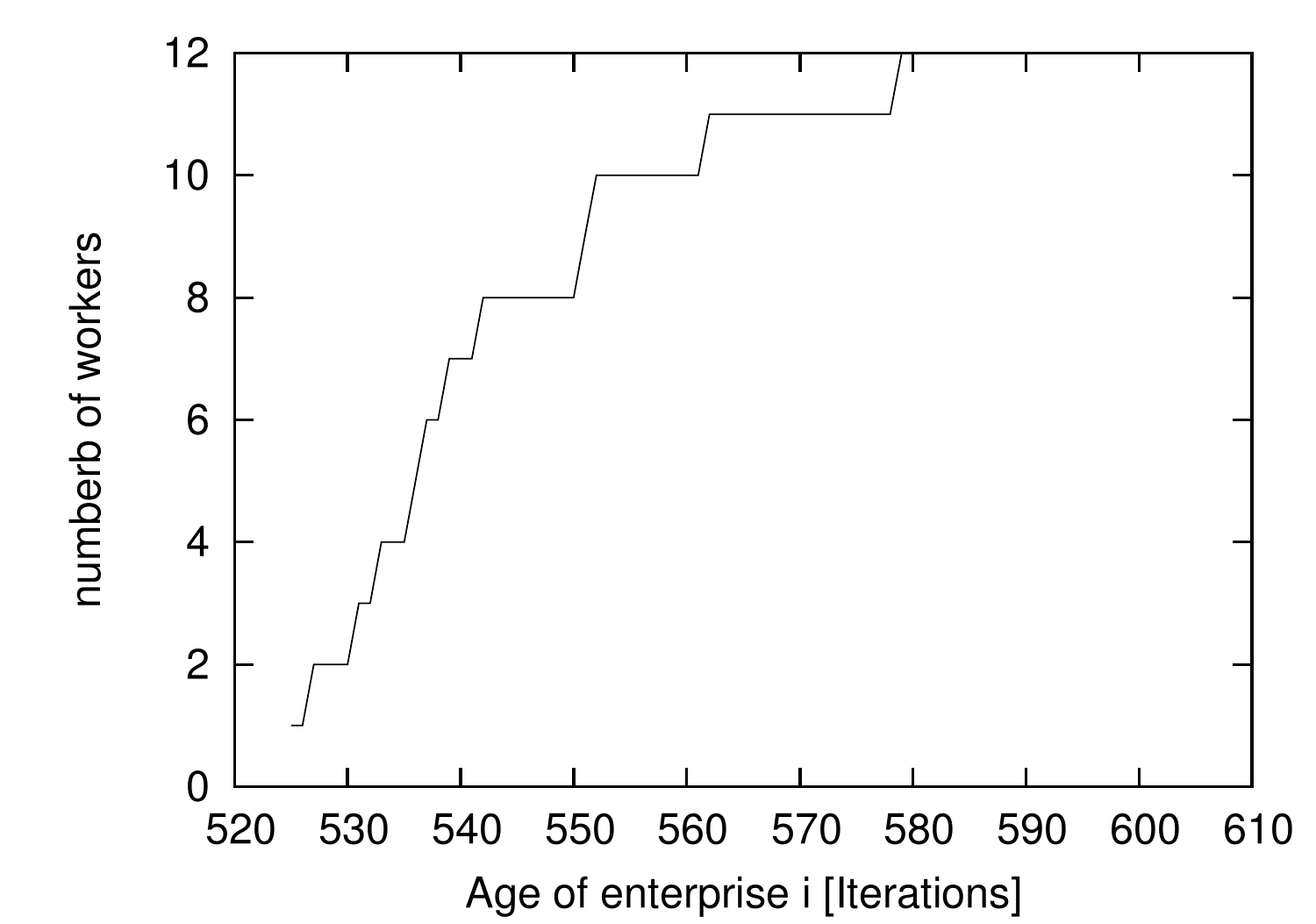}
\label{fig:333/example_hir24}
\caption{}
\end{subfigure}
\begin{subfigure}[b]{0.45\textwidth}
\includegraphics[angle=270, width =\textwidth]{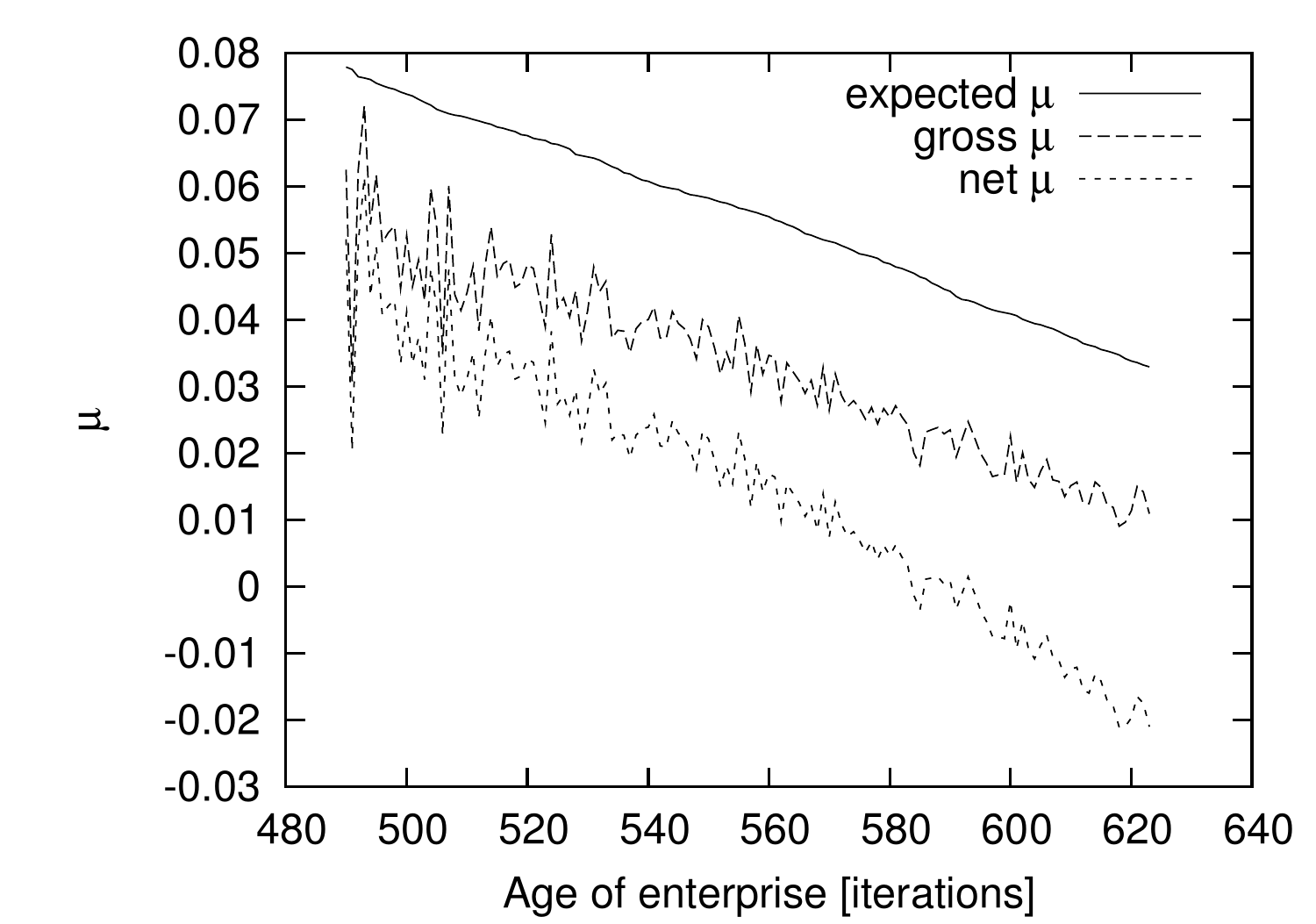}
\label{fig:333/example17}
\caption{}
\end{subfigure}
\begin{subfigure}[b]{0.45\textwidth}
\includegraphics[angle=270, width =\textwidth]{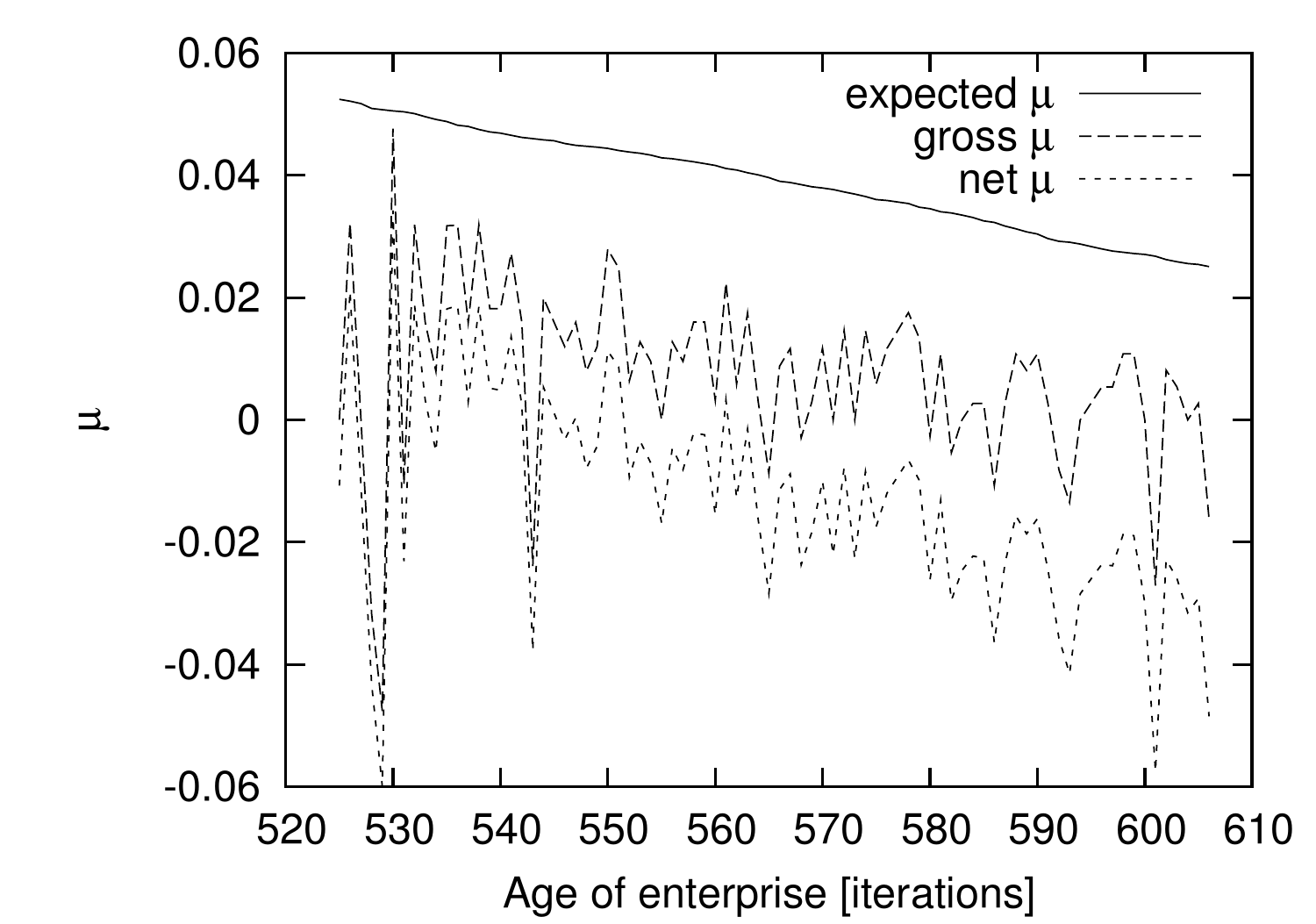}
\label{fig:333/example24}
\caption{}
\end{subfigure}
\caption{\small{Two examples of enterprises. Their sizes (upper line) and their margin and net margin (lower line) are plotted over their lifetime. Both arise from the same simulation with the same  parameters as in figure \ref{fig:sim330b} (interest rate $r=$ 0.075, $\nu= 8$, 450 active enterprises, 100000 workers, uniform margin distribution between 0 and 0.1, bankruptcy threshold $\gamma$= 2). The enterprise shown in figures (a) and (c) starts with a higher $\mu_i$ and grows much bigger, and exits for more iterations than the one shown in (b) and (d).}}
\label{fig:trajectories_mu}
\end{figure}

\begin{figure}[h!]
	\centering

\begin{subfigure}[b]{0.45\textwidth}
\includegraphics[angle=270, width =\textwidth]{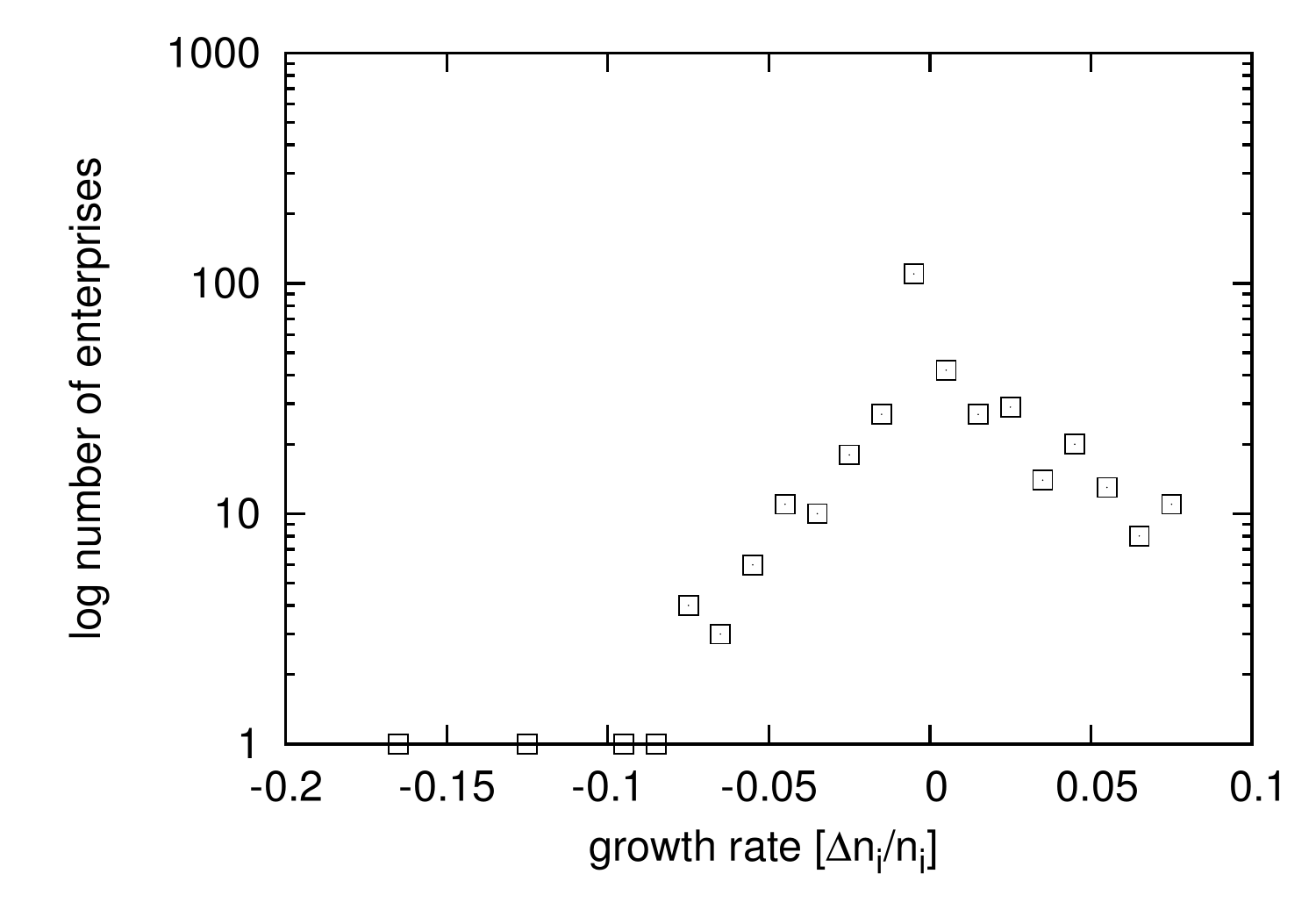}
\label{fig:335/log_growth_rate}
\caption{}
\end{subfigure}
\begin{subfigure}[b]{0.45\textwidth}
\includegraphics[angle=270, width =\textwidth]{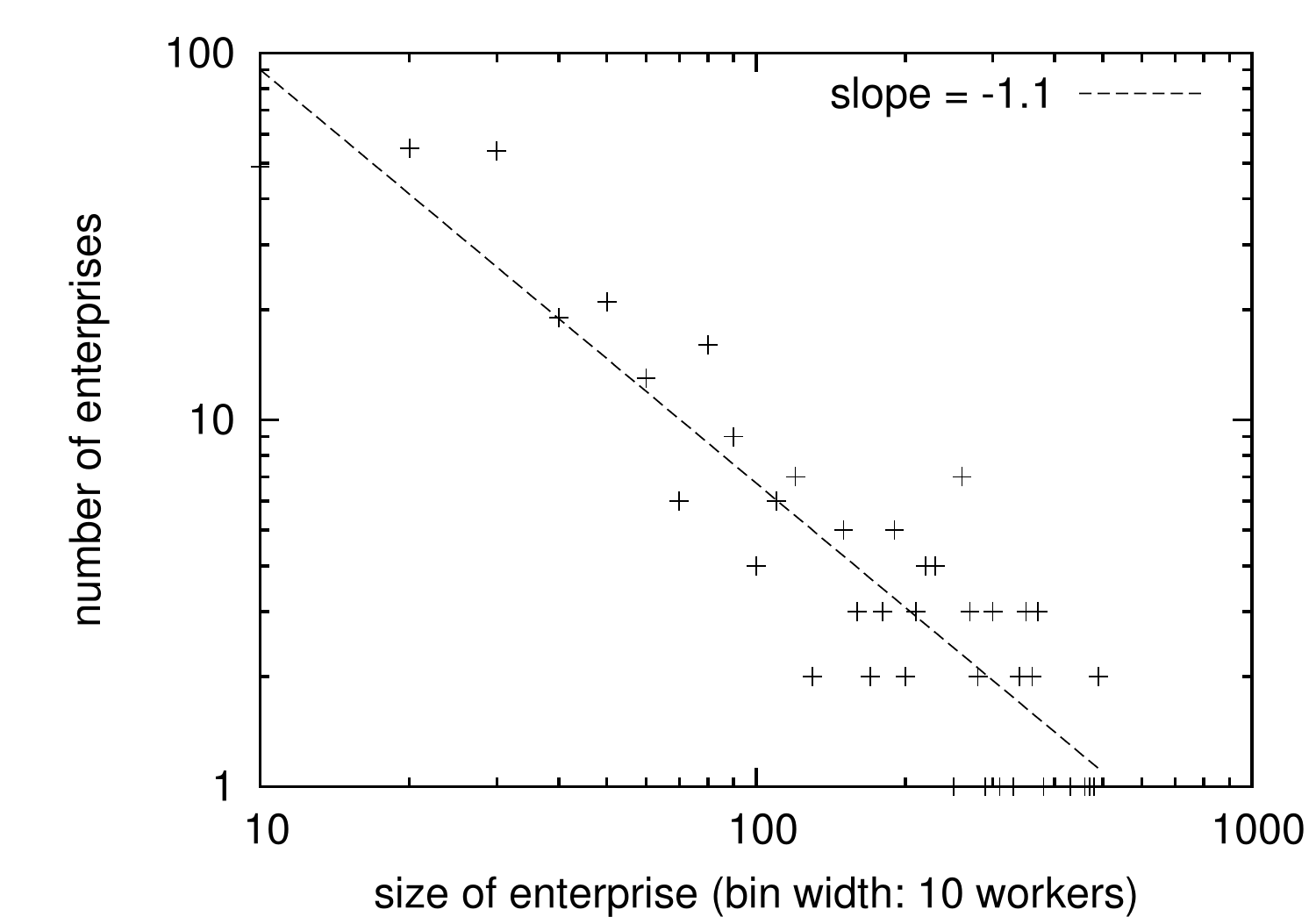}
\label{fig:335/log_growth_rate}
\caption{}
\end{subfigure}

\begin{subfigure}[b]{0.45\textwidth}
\includegraphics[angle=270, width =\textwidth]{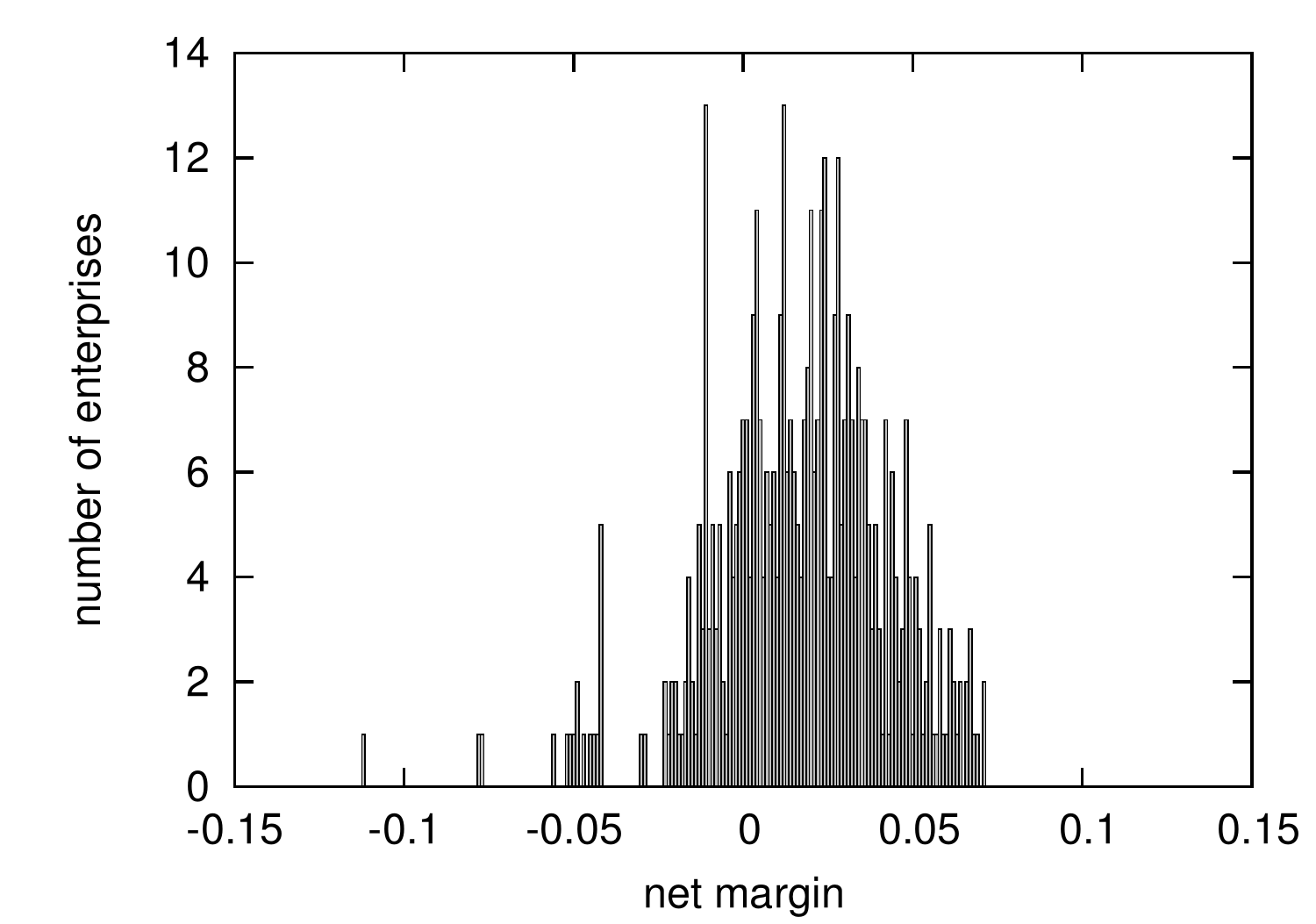}
\label{fig:335/number_ent_margin}
\caption{}
\end{subfigure}
\begin{subfigure}[b]{0.45\textwidth}
\includegraphics[angle=270, width =\textwidth]{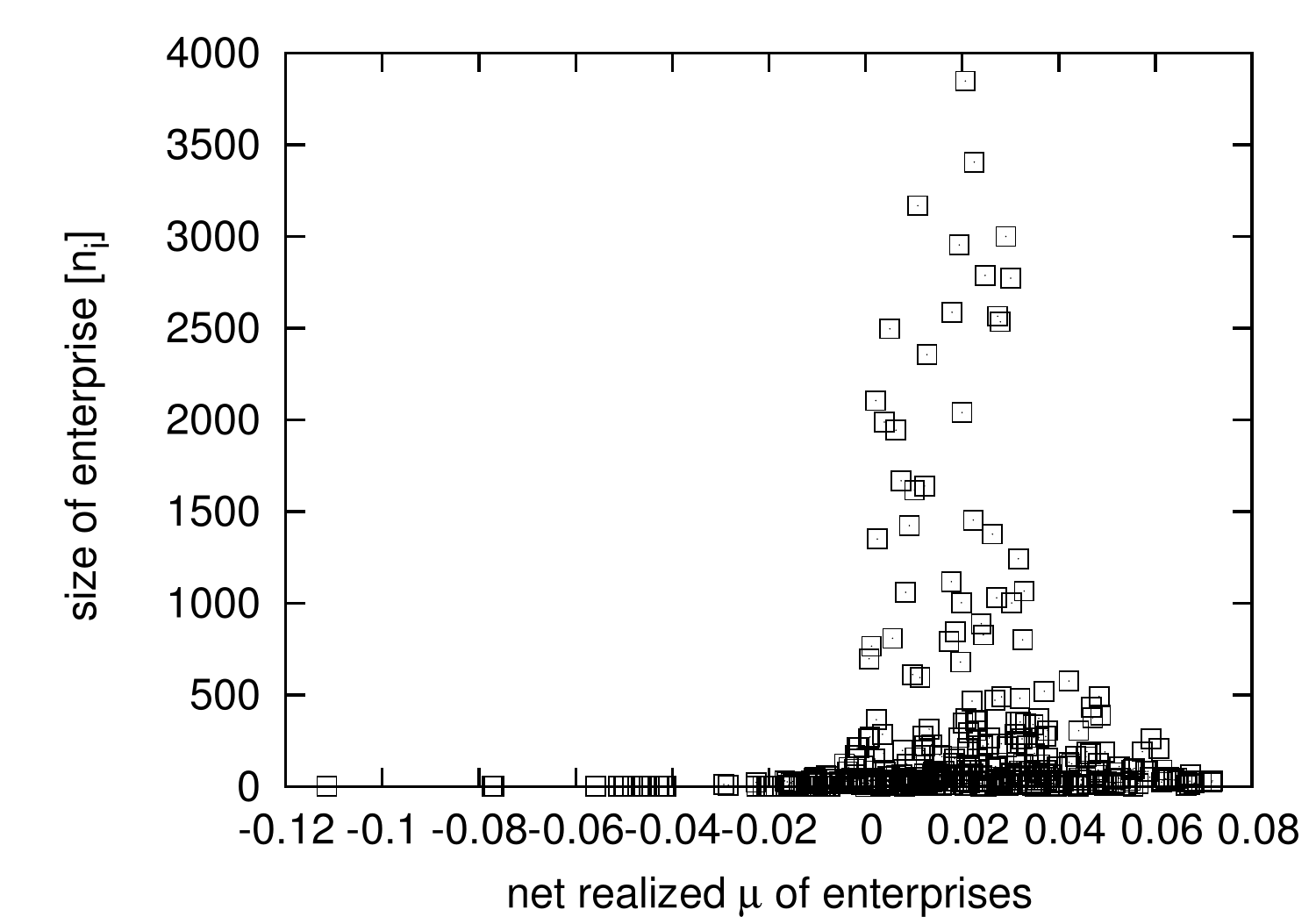}
\label{fig:335/margin_size}
\caption{}
\end{subfigure}
%\label{fig:growth_rate_335}
\caption{Interest rate 0.075, $\nu =4$, $\gamma =2$, 100000 workers, $\approx$ 450 active enterprises, unemployment is 0. Snapshots after 750 iterations. (a) growth rate distribution of enterprises, (b) size distribution of enterprises, (c) number of enterprises per net realized margin, (d) size of enterprises vs. net realized margin (every point represents one enterprise).
 If enterprises did not face a random constraint due to limited aggregate demand, every enterprise would grow according to its net realized margin, and the growth rate distribution (a) would have precisely the shape of the number of enterprises per net realized margin (c), which is clearly not the case. (d) shows that the dynamics of interest payments, retstarts, bankruptcies and recentering of the margin causes big enterprises to appear towards the middle of the net margin distribution, although it is noisy.}
% actually, the last two images are without the goods constraint! 

\label{fig:growth_rate_335}
\end{figure}

The lifetime of enterprises is limited by the centering of margins (which always decreases the margins of existing firms) and by the interest payments, which both may lead to bankruptcies. Figure (\ref{fig:trajectories_mu}) illustrates the time evolution of two particular enterprises of one simulated system. The left hand side figures (a) and (c) correspond to an enterprise with large initial $\mu_i$: $\mu_i= 0.078$. Created at iteration $t=490$, with $n_{i,0}=1$, it grows during about $100$ iterations (figure a) before starting to decline. It goes bankrupt after 134 periods. The bottom figure shows the evolution of its margins: centered expected gross margin $\mu_{i,t}$, actual gross margin (\ref{eq:mugross}) and actual net margin (\ref{eq:munet}). All the three margins decrease over time, but the net margin decreases faster, reflecting the interest burden due to interests owed for cumulated loans, which the firm has to pay, up to its eventual bankruptcy. The fluctuations in the gross and net realized margins reflect the randomness in the goods market. 
For a given bankruptcy threshold $\gamma$, the higher the interest rate, the shorter the typical lifetime of enterprises. This has implications on the size and growth rate distribution, which we trace at every iteration.
\paragraph{Size distribution of enterprises.}
Even though every single enterprise size varies over time as the typical cases shown in figure \ref{fig:trajectories_mu}, the size distribution of enterprises forms a fat tail distribution that can be approximated by a power law which remains stable (unlike the case without new enterprises and bankruptcies where the steady state is a monopoly). The stochastic process leading to this is a superposition of two effects: a stochastic process due to competition for demand (on its own described in \cite{MetzigGordon2012Physica}), and the dynamics that each enterprise grows (and later shrinks) over its lifetime. Depending on $r$ and the number of new enterprises per iteration, the numerically found exponent of the approximate power law varies between 0.8 and 2. %  Analysis of the distribution of the net margin of all enterprises, and the net margin where workers are hired explains this:
 The higher the interest rate and the number of entries per iteration, the shorter the average lifetime of an enterprise. Only for low interest rates, this `turnover' is slow enough that certain enterprises have enough time to reach large sizes, before their renormalized margin makes them less competitive compared to younger enterprises. For high interest rates the distribution is steeper, i.e. the power law exponent is larger. An example with a relatively low interest rate is given in figure \ref{fig:growth_rate_335}.
%\textcolor{red}{Analysis of the time evolution of the enterprises' size vs margin distribution shows this.} 
% This growth process due to their margin is superposed by the random process described in section \ref{sec:proba} and in \cite{MetzigGordon2012Physica}, since in all the simulations enterprises are in competition for aggregate demand or workforce: if enterprises spend their expected net profits, and workers their salary, the aggregate demand will be inferior to the aggregate offer, because a flow of money goes to the bank for interest payments. Therefore, the explanation in \cite{MetzigGordon2012Physica} holds also in this more complicated setting, as figure \ref{fig:growth_rate_335} shows.
\paragraph{Growth rate distribution of enterprises.}
For a simpler scenario without margin heterogeneity, the model provides an explanation for the stylized fact of a tent-shaped growth rate distribution, both analytically and in simulations \cite{MetzigGordon2012Physica}. The necessary ingredient is that enterprises are in competition for a limited quantity (workforce or aggregate demand).
%Implications are that big enterprises contribute more to the peak of the tent-shaped growth rate distribution, and small enterprises are the reason for its fat tails.
Big enterprises add to the peak of the tent-shaped growth rate distribution, and the small contribute more to the tails. In the setting of this paper, the growth rate of an enterprise depends (in addition to the randomly attributed limited sales opportunities) on its expected gross margin via equation (\ref{eq:q_p_i}). Margin heterogeneity on its own can distort the tent-shaped growth rate distribution. If however the combination of (i) interest payments, (ii) entry and exit of enterprises, and (iii) recentering of the margin produces a situation where the big enterprises have a realized net margin that is in the middle of the distribution, we find a tent-shape growth rate distribution. To that aim it is necessary that the growth dynamics produces a correlation between the average growth rate and size of an enterprise. The elements (i) -- (iii) provide this, as is illustrated in figure \ref{fig:growth_rate_335}. The correlation is the following: the big enterprises cumulate in the middle of the distribution of the net realized margin. They rarely have very high net realized margins, since it takes them time to grow, and over time their expected gross margin $\mu_i$ declines. The biggest enterprises are also less represented among the enterprises with the lowest net realized margins, since these have already lowered their production (according to equation \ref{eq:q_p_i}).

%start to lower their production and offer less jobs if their net realized profits are negative. %(The latter is influenced by the bankruptcy threshold $\gamma$). %Theoretically, the big enterprises add to the peak of the tent-shaped growth rate distribution, so if the described growth dynamics produces indeed a situation where the big enterprises have a realized net margin that is in the middle of the distribution, we find indeed a tent-shape growth rate distribution, as is illustrated in figure \ref{fig:growth_rate_335}. 
The small enterprises are spread all over the distribution of the net margin, so they contribute to the fat tails of the tent-shaped distribution. Subfigure (d) in figure \ref{fig:growth_rate_335} shows that it must be indeed small enterprises who have a net realized margin far from the center of the distribution.
It may be noisy due to the randomness in the system.
%Another factor that limits the effect of margin heterogeneity on the growth rate is that the number of enterprises per net margin has a maximum towards the middle of the distribution. see subfigure (c) in figure \ref{fig:growth_rate_335}).

A case where the tent-shaped growth rate distribution is extremely noisy or nonexistent is when big enterprises exist with net margins far from the center of the distribution. This can arise from fluctuations in the initialisation of new enterprises, or from bankruptcy waves. If the competition among enterprises is removed, e.g. by deficit spending of the bank in combination with temporal unemployment, the probabilistic dynamics that generated the tent-shaped growth rate distribution is also removed. In that case, the growth rate distribution adopts, as expected, the same shape as the number of enterprises per net margin (subfigure (c) in figure \ref{fig:growth_rate_335}). The distributions in subfigure (c) and (d) are centered around a positive value, whereas the growth rate distribution in (a) is centered around zero, because unemployment is zero, and enterprises face a constraints both in aggregate demand and in the available workforce. Despite having realized positive net profits they do not grow on average, because not enough workforce is available.

\newpage

\subsection{Fluctuations}\label{sec:fluct}
\begin{figure}[h!]
	\centering
\begin{subfigure}[b]{0.45\textwidth}
\includegraphics[angle=270, width =\textwidth]{unemployment330b.pdf}
%\label{fig:33od/unemployment_b}
\caption{}
\end{subfigure}
\begin{subfigure}[b]{0.45\textwidth}
\includegraphics[angle=270, width =\textwidth]{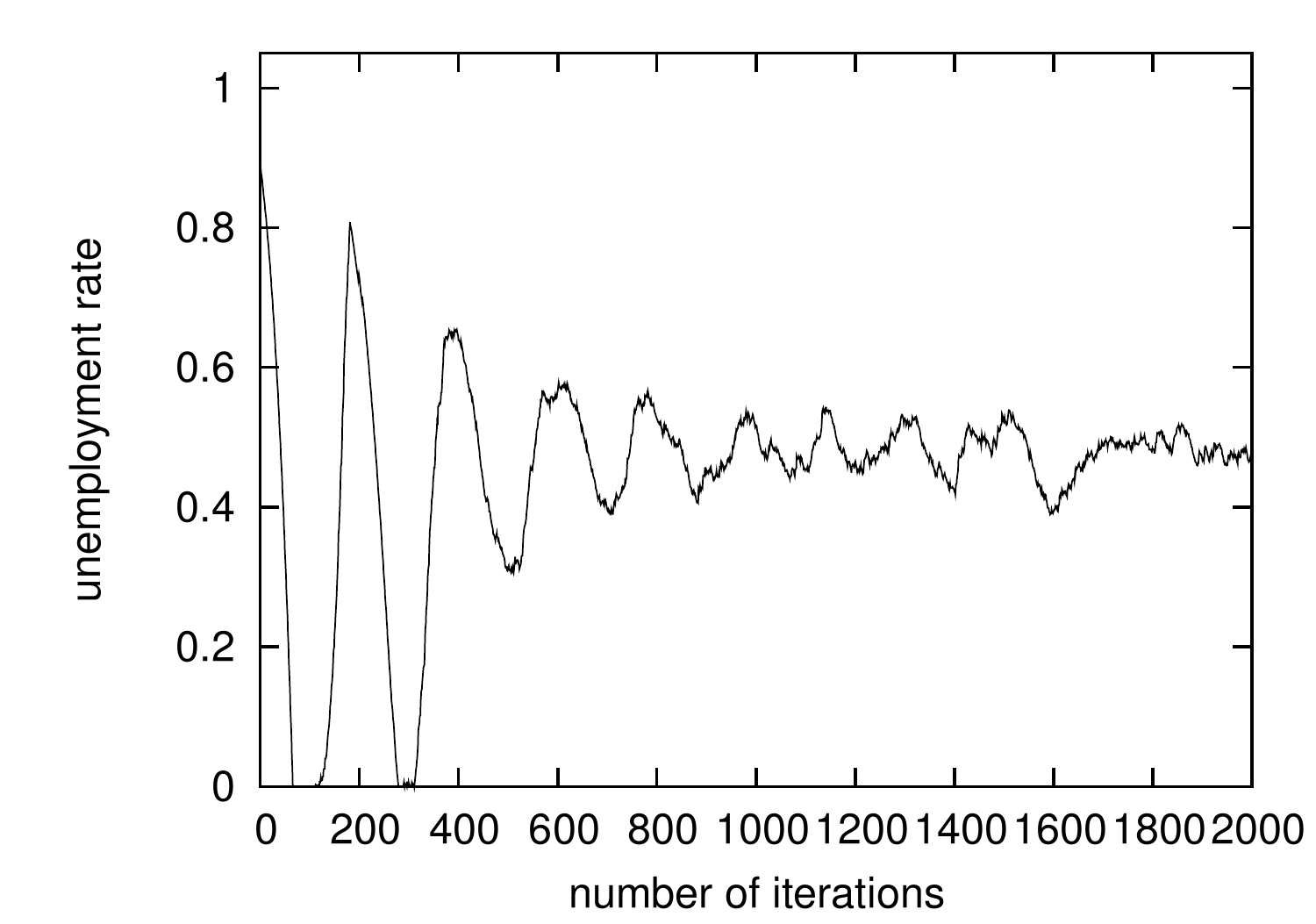}
%\label{fig:33od/unemployment_d}
\caption{}
\end{subfigure}
 	\caption{Interest rate $r= 0.011$, $\nu =8$, $\gamma =2$, $N_w = 100000$. Time evolution of unemployment for two different widths of the $\mu$-distribution. In figure (a), $\mu_i$ for new enterprises are drawn randomly between 0 and 0.1 (as in the rest of the paper), in figure (b), between 0.025 and 0.075. Two effects are visible: (i) Unemployment clearly fluctuates more in the system with a larger width of the $\mu$-distribution. In figure (b), all entering enterprises have more similar growth behaviour and a more similar lifetime, so unemployment fluctuates less. (ii) Unemployment is higher in (b), since the highest margin is only 0.075, so compared to the scenario in (a) a smaller fraction of enterprises has a net realized margin bigger than zero, meaning that only few enterprises will grow and offer more posts.}
\label{fig:sim330dunemployment}
\end{figure}

Stability has been verified by simulating 10 times the system with same parameters and different randomization, as well as by simulating the same system with different initial conditions. With respect to that, the system is stable, though atypical cases may in principle arise. For the observed simulations, after some time fluctuations occur always around the same mean, which is characterized by the parameters $\nu$, $\gamma$, $r$ and the range of $\mu_i$. The higher the interest rate and the rate of new entries per iteration, the faster the system loses the memory of the initial conditions, and the fluctuations due to the randomness in the system become dominant quickly. 
Depending on the width of the distribution of $\mu$, the observed fluctuations can be strong. %sim330d, sim330e
A rise in unemployment is caused by some big enterprise going bankrupt. This can be seen in figures \ref{fig:sim330b} and \ref{fig:sim330dunemployment} wherever unemployment rises very fast. Successively, it may rise even further, because other enterprises lower their production due to a big lack of demand. Once unemployment starts to decline, this happens slower than it rose, because it takes time for other enterprises to grow and absorb the workforce.

Roughly, this is the type of fluctuations that DelliGatti et al. \cite{DelliGatti2011} interpret as business fluctuations in their similar model. Since the evolution of the job losses due to bankruptcy is not explicitely given in their book, we estimate its impact on unemployment to be at least twice as strong as in the case presented here, based on their bankruptcy frequency and the relation of enterprises to workers.

\paragraph{Role of Bankruptcy threshold on fluctuations.}
%sim330e
The lower $\gamma$, (i.e. the less debts per hired worker enterprises can incur before going bankrupt), the less the system fluctuates. For low $\gamma$, enterprises go bankrupt while still growing, at a point of time when their average age and size are still small. The job losses due to bankruptcy are more evenly spread, and fluctuations have a lower amplitude and a higher frequency. Workers can be employed by other enterprises faster, resulting in lower unemployment. Not surprisingly, the lower the bankruptcy threshold, the lower the aggregate debt of enterprises.
%\paragraph{Comparison to the model in \cite{DelliGatti2011}.}
%\textcolor{red}{Do you find this too negative? I could compare even more aspects, but I am not sure if one learns much from it, because I cannot draw conclusions from the comparisons.} 

\section{Validation of the model}\label{sec:validation}
The strength of this model is that it can be compared both to a theory of a firm, as well as give rise to macroeconomic quantities. %On the firm level, in the setting presented in this paper, it is possible to analyze size, growth rate, age, and level of debt. 
On the macroeconomic level, it is possible to analyze distribution of the size, growth rate, age and debt, as well as aggregate quantities: unemployment, bankruptcy frequency, productivity increase (via the part by which $\mu_{eff}$ is recentered).
The tent-shaped growth rate distribution has been widely acknowledged as a stylized fact \cite{StanleyAmaraletal1996}, \cite{Botazzi2006}. %In our case, the tent-shape can be approximated by a $1/|g|$-distribution, but given the small range of growth rates, empirical evidence is insufficient to distiguish its validity e.g. from a Laplacian growth rate distribution. 
For firm size distribution, several authors equally propose a Zipf law \cite{Axtell2001}, \cite{OkuyamaTakayasu1999}, \cite{DelliGatti2008}, and the fat-tailed nature of the distribution is generally accepted.
The statistics about debt and age at failure (not presented in this paper) are qualitatively in agreement with data from the German Statistical Office, and the conclusions by \cite{ThornhillAmit2003}. 

\paragraph{}%{Effect of bankruptcies on the growth rate distribution of firms.}
The model offers also the possibility to compare tendencies with empirical surveys. For instance, as A. Coad states in his survey of the literature on firm growth \cite{Coad2007} p.12, ``there is a lot of evidence that a slight negative dependence of the growth rate on size is present at various levels of industrial aggregation.'' %\textcolor{red}{cite here Sutton who says the same!} 
This effect also occurs in our model: To reach a large size, enterprises need to attain a certain age. During that time, their margin diminishes due to the recentering of $\mu_{eff}$, and the level of debt will increase, so it is not possible for the net margin of big enterprises to range among the higher end of the distribution. Small enterprises, in turn, are more likely to have a net margin above average: Of the enterprises with negative net margin, a part goes bankrupt, and evidently cannot enter the statistics on average growth rates, so the average growth rate is shifted towards a positive value of growth. %It depends on how initial debt and interest rate are initialized how high the average growth rate of small enterprises is. This could be extracted from data. %\textcolor{red}{Is there some quantitative analysis among the cited ones p. 12 survey?}
A second tendency produced by the model has empirically been found by \cite{StanleyAmaraletal1996}: The width of the tent-shape of the growth rate distribution depends on the enterprise sizes that are considered. If only the big enterprises are considered, the distribution is much narrower than if it is computed for a sample of small enterprises. However, a certain range of sizes is necessary. If only enterprises of precisely the same size add to the growth rate distribution, our model would produce its shape to be Gaussian.  

The parameters in the model are all relative, i.e. there is no specific meaning what timespan is described by one iteration. Roughly one month could be the order of magnitude. The type of analysis in this paper is rather suited to understand dependencies, and to tackle the question how much randomness a system has or should have, with respect to the deterministic part of the dynamics. The relative weight of these two elements is visualized in trajectories of the expected and net $\mu$ shown in figure \ref{fig:trajectories_mu}. %For this methodology it is important to study both quantities per enterprise (such as growth rate, margin) as well as quantities per worker (such as job losses, effective margin, bankruptcy threhold).

\section{Discussion and Conclusion}
\label{sec:conclusion}

In this paper we proposed an agent-based model linking statistical regularities of macroeconomic systems to characteristics of single firms. This is done by combining a well understood simpler model in a forthcoming paper \cite{MetzigGordon2012Physica} with new features, in a way that preserves the mechanisms and thus the arising statistical regularities. The new elements are heterogeneous margins, interest payments, bankruptcies and recentering of the margin of firms, reflecting an `ageing process'. Altogether, the system exhibits a fat tail size distribution (which can be approximated by a power law), a tent-shape growth rate distribution as well as fluctuations in unemployment, in aggregate debt level and in the number of active enterprises. These fluctuations can be interpreted as business fluctuations. Other possible ways to describe business fluctuations with this model would be to vary some parameters that are currently fixed, such as the effective margin (reflecting the profit margins), the interest rate or the frequency of new entries of enterprises, in order to reproduce a specific situation observed in the real world. 

%In order to avoid over-interpretation of the results, we summarize the limitiations of the model, 
Finally we would like to point out some interesting problems for which this model can be useful: (1) More extended comparisons to data are in progress for growth rates and bankruptcies, but are possible for other quantities such as credit constraints.
 (2) In the model presented here, the margin of one enterprise is cannot be influenced by its own behaviour throughout an enterprise's lifetime. Productivity increase happens thus only via the creation of new enterprises. A possible extension of the model would be to let the margin of an enterprise depend on its investments, as is done in \cite{DelliGatti2011}, or by purchasing production goods that depreciate over time as \cite{Bruun2008}. This would allow for enterprises to take strategic choices and to have more variable lifecycles.  
(3) The financial sector is very simplified in the model. Decisions of the bank depending on its own balance sheet are not present, as well as variable interest rates or more than one bank, which seems a promising extension in order to study more in detail the role of credit.


\begin{thebibliography} {5}
%\bibitem{Gaffeo2008}Gaffeo, Edoardo, Delli Gatti, Domenico, Desiderio, Saul and Gallegati, Mauro, \emph{Adaptive Microfoundations for Emergent Macroeconomics}. Eastern Economic Journal, Vol. 34, Issue 4, pp. 441-463, (2008)
\bibitem{BruunHeynJohnsen2008} C. Bruun, C. Heyn-Johnsen \emph{The Paradox of Monetary Profits: An Obstacle to Understanding Financial and Economic Crises? }, Discussion Paper No. 2009-52, Economics-E-Journal (2009)
\bibitem{Bezemer2010} Bezemer, D. J. \emph{Understanding Financial Crisis Through Accounting Models}, Accounting, Organizations and Society, Vol 35 Issue 5, pp. 676 - 688, Elsevier (2010)
%\bibitem{Bezemer2011} Bezemer, D. J. \emph{Causes of Financial Instability: Don't Forget Finance}, Working Paper No. 665, Levy Economics Institute (2011) 
\bibitem{Seppecher2009} Seppecher, Pascal, \emph{Un Mod\`ele Macro\'economique Multi-Agents avec Monnaie Endog\`ene}, Document de Travail 2009-11, Groupement de Recherche en Economie Quantitative d'Aix-Marseille (GREQAM) (2009)
\bibitem{GodleyBook2007} Godley, W., Lavoie, M., \emph{Monetary Economics -- An integrated Approach to Credit, Money, Income, Production and Wealth}, Palgrave Macmillan, New York (2007)
\bibitem{Kalecki1942} Kalecki, M., \emph{The Determinants of Profits} (1942) in: Selected Essays on the Dynamics of the Capitalist Economy 1933 - 1970, Cambridge University Press (1971)
\bibitem{Keen2010} Keen, Steve, \emph{Solving the Paradox of Monetary Profits}, Discussion Paper No. 2010-2, Economics-E-Journal (2010)
\bibitem{Bruun2008} Bruun, Charlotte \emph{Rediscovering the Economics of Keynes in an Agent-Based Computational Setting}, Agent-Based Modelling in Economics and Finance, Trento, Italy (2008)
\bibitem{DelliGatti2008} Delli Gatti, D., Gaffeo, E. Gallegati, M., Giulioni, G., Palestrini, A., \emph{Emergent Macroeconomics}, Springer Italia (2008)
\bibitem{DelliGatti2011} Delli Gatti, D., Desiderio, S., Gaffeo, E., Cirillo, P., Gallegati, M., \emph{Macroeconomics from the Bottom-up}, Springer Italia (2011)
%\bibitem{Marsili1998} Marsili, M., Zhang, Y.-C., \emph{Interacting Individuals Leading to Zipf's Law}, Phys. Rev. Lett. 80, pp. 2741 - 2747 (1998) 
%\bibitem{Epstein1999} Epstein, J. M., \emph{Agent-Based Computational Models and Generative Social Science}, Complexity, Vol. 4 No 5, John Wiley and Sons (1999)
\bibitem{Botazzi2006} Botazzi, G., Secchi, A., \emph{Explaining the distribution of firm growth rates}, The Rand Journal of Economics, 37, 2 p. 235 (2006) 
\bibitem{Sutton1997} Sutton, J., \emph{Gibrat's Legacy}, Journal of Economic Literature, Vol. XXXV, pp. 40 - 59 (1997)
\bibitem{ThornhillAmit2003} Thornhill, S., Amit, R., \emph{Learning about Failure: Bankruptcy, Firm Age, and the Resource-Based View}, Organization Science, Vol. 14 nb. 5, pp. 497-509 (2003)
%\bibitem{MalevergneSaichevSornette2008} Malevergne, Y., Saichev, A., Sornette, D., \emph{Zipf's law for firms: relevance of birth and death processes},  Available at SSRN: http:\/\/ssrn.com\/abstract=1083962 or http:\/\/dx.doi.org\/10.2139\/ssrn.1083962 (2008)
\bibitem{Coad2007} Coad, A., {Firm Growth: A Survey} Papers on Economics and Evolution 2007-03, Max Planck Institute of Economics, Evolutionary Economics Group (2007)
%\bibitem{HymerPashigian1962} Hymer, S., Pashigian, O., \emph{Firm Size and Rate of Growth}, Journal of Political Economy 70 (6), pp. 556 - 569 (1962)
%\bibitem{Bezemer2012} Bezemer, D. J., \emph{Finance and Growth: When Credit helps, and When it Hinders}, Conference Paper, Plenary Conference of the Institute of New Economic Thinking, Berlin (2012)
%\bibitem{West1999} West, G. B., \emph{The Origin of Universal Scaling Laws in Biology}, Physica A nb 263, pp. 104 - 113 (1999)
%\bibitem{Gibrat1931} Gibrat, R., \emph{Les in\'egalit\'es \'economiques}, PhD thesis, Librairies du Receuil Sirey, Paris (1931)
%\bibitem{Takayasu1997} Takayasu, H., Sato, A.-H., Takayasu, M., \emph{Stable Infinite Variance Fluctuations in Randomly Amplified Langevin Systems}, Phys. Rev. Lett. 79, pp. 966 - 969 (1997)
%\bibitem{Sornette1998} Sornette, D., \emph{Multiplicative processes and power laws}, Phys. Rev. E 57, pp. 4811 - 4813 (1998)
%\bibitem{Zanette1997} Zanette, D. H., Manrubia, S. C., \emph{Role of Intermittency in Urban Development: A Model of Large-Scale City Formation}, Phys. Rev. Lett. 79, pp. 523 - 526 (1997)
\bibitem{BlankSolomon2000} Blank, A., Solomon, S., \emph{Power laws in cities population, financial markets and internet sites (scaling in systems with a variable number of components)}, Physica A 287, pp. 279 - 288 (2000) 
%\bibitem{BiroJakovac2005} Bir\'o, T. S., Jakov\'ac, A., \emph{Power-law tails from multiplicative noise}, Phys. Rev. Lett. 94, pp. 132302 - 132305 (2005)
\bibitem{MetzigGordon2012Physica} Metzig, C., Gordon, M. B., \emph{} (2012, submitted)
\bibitem{Axtell2001} Axtell, R. L., \emph{Zipf Distribution of U.S. Firm Sizes}, Science Vol. 293 no. 5536 pp. 1818-1820 (2001)
%\bibitem{SaichevMalevergneSornette2011} Saichev, A., Malevergne, Y., Sornette, D., \emph{Theory of Zipf's Law and Beyond}, Springer (2009)
%\bibitem{RichmondSolomon2001}Richmond, P., Solomon, S., \emph{Power Laws are disguised Boltzmann Laws}, Int. Journal Mod. Phys. C, Vol 12 No. 3 pp. 333 - 343 (2001)
\bibitem{StanleyAmaraletal1996}Stanley, M. H. R., Amaral, L. A. N., Buldyrev, S. V., Havlin, S. Leschhorn, H., Maass, Ph., Salinger, M. A., Stanley, H. E., \emph{Scaling behaviour in the growth of companies}, Nature, Vol. 379 (1996), pp. 804-806 
\bibitem{OkuyamaTakayasu1999}Okuyama, K. Takayasu, M., Takayasu, H., \emph{Zipf's law in income distribution of companies}, Physica A 269 pp. 125 - 131 (1999)
\end{thebibliography}
\end{document}